\documentclass[letterpaper]{article}

\usepackage{aaai2027}

\usepackage{helvet}
\usepackage{courier}
\usepackage[hyphens]{url}
\usepackage{graphicx}
\usepackage{natbib}
\usepackage{caption}
\usepackage{xcolor}
\usepackage{booktabs}
\usepackage{array}
\usepackage{multirow}
\usepackage{amsmath}
\usepackage{amssymb}
\usepackage{algpseudocode}
\usepackage{placeins}

\newcounter{windalgorithm}
\newenvironment{windalgorithm}[2]{%
    \refstepcounter{windalgorithm}%
    \par\medskip\noindent\rule{\columnwidth}{0.4pt}\par
    \noindent\textbf{Algorithm~\thewindalgorithm: #1}\label{#2}\par\vspace{2pt}
}{%
    \par\noindent\rule{\columnwidth}{0.4pt}\par\medskip
}

\title{Guarded Agent-in-the-Loop Residual Adaptation for Evolving Wind Power Regimes}
\author{Yuhang Dong, Min Xia, Jinghong Li, Yue Dai, Xinyue Wang, Shi Liang\corresponding}
\affiliations{Nanjing University of Information Science and Technology, Nanjing, China}

\begin{document}
\maketitle

\begin{abstract}
Wind power forecasting models are often deployed as fixed temporal backbones, but the errors they make can vary across meteorological regimes such as cold waves, heat-and-calm periods, productive typhoon, and destructive storms. We present a guarded Agent-in-the-loop residual adaptation framework that leaves the temporal backbone unchanged and adapts only a small set of scene-conditioned residual experts. An Agent layer interprets external scene information, identifies the current regime, selects residual experts, and applies a correction only when recent observed errors support it. New experts enter the pool after they outperform their initializations on a validation set, keeping model growth under explicit control. Backbone-only and Agent-enhanced forecasts are compared at lead times of 4, 8, 12, 24, and 48 across nine time-series backbones. Across these runs, the guarded framework lowers both MAE and RMSE for all 45 backbone-horizon combinations, with average MAE reductions ranging from 11.21\% to 20.26\% across backbones. Online-selection baselines, component ablations, and routing analyses identify the roles of residual experts, scene-aware routing, and the online acceptance rule. Candidate-scene runs evaluate the promotion path under controlled conditions; long-term unattended deployment was not evaluated.
\end{abstract}

\section{Introduction}

Short-term wind-power forecasting is increasingly handled by neural time-series backbones that map historical power, turbine signals, and weather inputs to future generation. Temporal modelling has improved, but deployment remains largely static. After training, the same mapping is used across normal operation, cold waves, heat-and-calm periods, productive typhoon, and destructive storm conditions. Input distribution shift alone does not describe the deployment problem. The same regimes can also reshape the residual error left by an otherwise strong temporal model.

For residual errors of this kind, adaptation is better framed as residual control than as backbone redesign. We keep the backbone as the temporal modeller and place an Agent layer around its residual path. Its authority is limited: it reads external scene evidence, forms a scene belief, routes the base forecast through residual experts, and accepts the correction only when observed feedback justifies deployment. Forecasting is then treated as communication between a temporal backbone and scene-reasoning Agents, not as one end-to-end mapping that absorbs every source of context.

The separation creates three practical difficulties. First, scene cues arrive as satellite quicklooks, field reports, real-time weather, event records, and turbine measurements, so they must be aligned without file names, folder labels, or future observations. Second, a scene-aware correction helps only when the routing Agent selects useful residual experts instead of collapsing to one expert or ignoring scene evidence. Third, deployment needs an explicit control rule: a new correction should enter the forecasting loop only after recent errors support it and, for candidate scenes, after validation justifies pool expansion.

We test this guarded Agent-in-the-loop view with paired multi-horizon forecasts, with and without the Agent layer. Separate diagnostics check whether the gain passes through the specified residual pool, routing, and acceptance paths. Evidence for autonomous evolution would require independent wind farms, larger rare-event support, and strict event-disjoint future evaluation.

\noindent\textbf{Contributions.}
\begin{itemize}
  \item A guarded Agent-in-the-loop residual formulation separates temporal modelling, scene reasoning, and correction in wind-power forecasting.
  \item Scene-conditioned residual routing is coupled with an online acceptance guard, tying expert use to observed forecast feedback.
  \item Experiments cover nine backbones at leads 4, 8, 12, 24, and 48, with audits for residual experts, routing behaviour, online acceptance, and candidate-scene promotion.
\end{itemize}

\section{Related Work}

Wind-power forecasting research has mainly improved either the temporal predictor or the environmental information supplied to it \citep{wang2021wind}. Existing work builds on attention-based sequence modelling, efficient Transformers, recurrent probabilistic forecasting, interpretable multi-horizon networks, and neural basis expansion models \citep{vaswani2017attention,kitaev2020reformer,lai2018modeling,salinas2020deepar,oreshkin2020nbeats,lim2021tft}. Recent long-sequence models further use sparse attention, decomposition, frequency modelling, patch tokens, temporal variation modelling, and inverted channel representations to strengthen the backbone \citep{zhou2021informer,wu2021autoformer,zhou2022fedformer,zeng2023transformers,nie2023patchtst,wu2023timesnet,liu2024itransformer}. In this study, these models are treated as reusable temporal engines. The central question is whether an auditable Agent layer can improve the same backbone while acting only on the residual path.

Conditional computation and lightweight adaptation provide the closest mechanistic vocabulary for this residual path. Mixture-of-experts methods route inputs to specialised parameters \citep{jacobs1991adaptive,shazeer2017moe}, while LoRA-style updates show that useful adaptation need not modify every parameter \citep{hu2022lora}. The guarded deployment view is also connected to online expert weighting, dynamic model averaging, selective prediction, and probabilistic forecast evaluation \citep{freund1997online,raftery2010dma,geifman2017selective,gneiting2007probabilistic}. Those methods are useful, but their target differs from ours. Here, the selected objects are post-backbone residual experts, and each comparison keeps the forecasting backbone fixed. Our comparison concerns guarded residual control, not a separate MoE or adapter benchmark.

Continual learning studies how systems retain and expand capability over time \citep{parisi2019continual,kirkpatrick2017overcoming,rebuffi2017icarl}. Wind-power forecasting adds an operational constraint: rare-event evidence is sparse, and future leakage is easy to introduce. Our Agent-in-the-loop design splits online adaptation into explicit responsibilities. Evidence Agents align scene cues, scene Agents form an auditable belief, routing Agents select residual experts, and candidate-scene Agents validate updates before registration. Multimodal encoders and language-model agents supply useful interfaces for this evidence layer, while the numerical forecaster remains a separate component \citep{radford2021clip,alayrac2022flamingo,yao2023react,shinn2023reflexion,wang2023voyager}.

\section{Guarded Agent-in-the-loop Method}

\begin{figure*}[t]
  \centering
  \includegraphics[width=\textwidth,height=0.65\textheight]{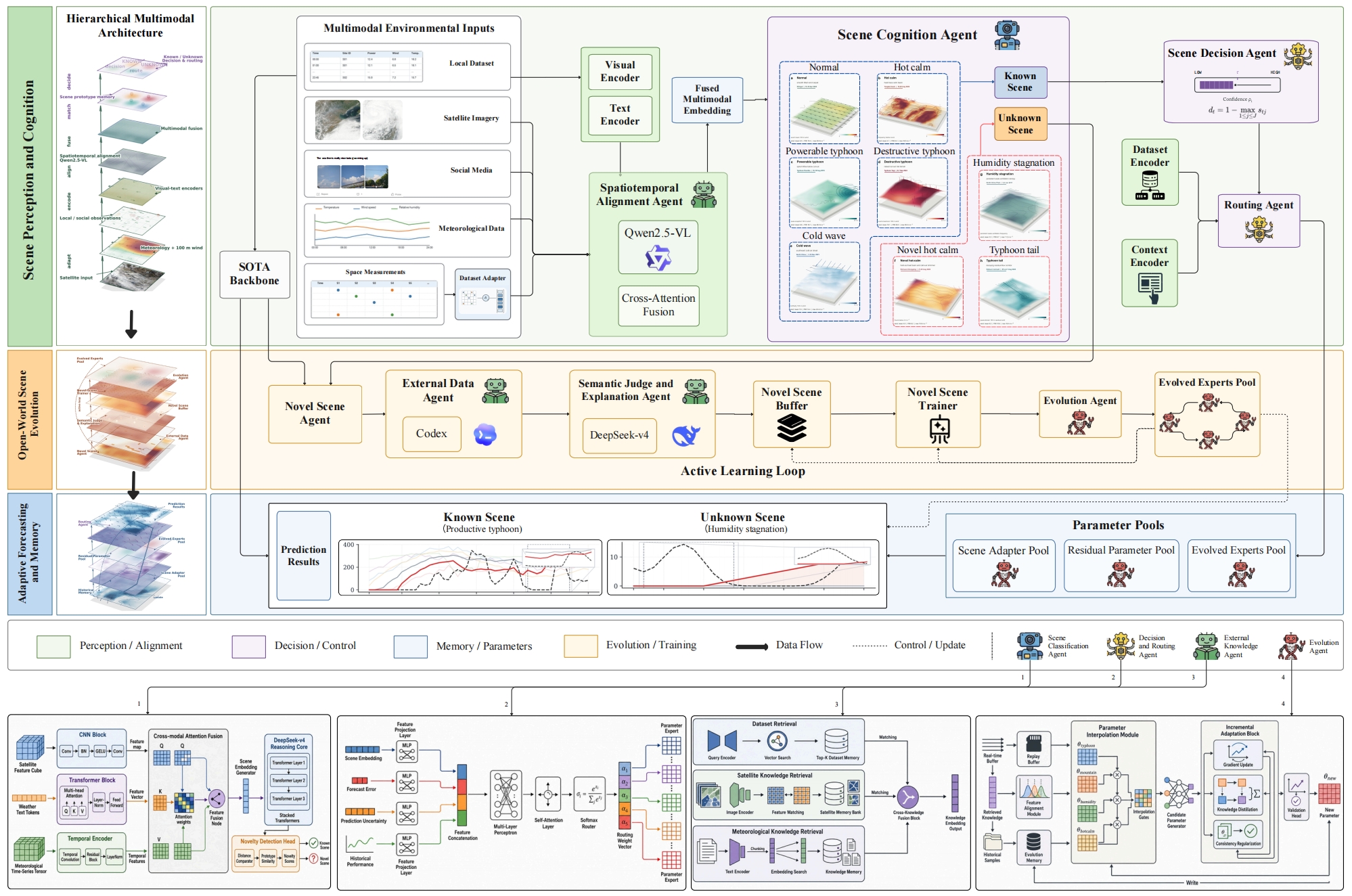}
  \caption{Guarded Agent-in-the-loop residual-adaptation framework: the backbone models temporal dynamics, while Agents organize scene evidence, route residual experts, guard corrections, and audit candidate-scene updates.}
  \label{fig:framework}
\end{figure*}

Figure~\ref{fig:framework} shows the information flow used throughout the method. Heterogeneous scene evidence is first converted into a causal input packet and then passed through an Agent communication path to obtain a scene representation. To keep evolution auditable, Agent actions are limited to residual expert routing and validation-gated pool updates. The forecasting backbone remains the temporal model, while the Agent layer controls how scene evidence modifies the residual correction.

\subsection{Problem Formulation}

Let $t$ be the forecast issue time and let $H$ be the forecast horizon. The causal input packet is $X_t=\{p_t,m_t,v_t,e_t\}$, where $p_t$ is the historical power and SCADA vector, $m_t$ is meteorology, $v_t$ is visual or field evidence, and $e_t$ is event context. Encoding maps the packet to a scene state $z_t$, after which the Agent layer chooses an action $a_t$ from known-scene routing or candidate-scene evaluation:
\begin{equation}
 z_t=\operatorname{ENCODE}(X_t),\qquad
 a_t=\operatorname{DECIDE}(z_t,M_{t-1}).
 \label{eq:state}
\end{equation}
Here $M_{t-1}$ is the scene and expert memory before time $t$. A backbone with parameters $\theta_B$ produces a base forecast $\hat{y}^{(0)}_{t+1:t+H}$. Keeping this backbone intact, the Agent layer selects a residual expert path:
\begin{equation}
 \hat{y}_{t+1:t+H}=\operatorname{BACKBONE}_{\theta_B}(p_t,m_t)+r_t.
 \label{eq:forecast}
\end{equation}
Our design is based on the assumption that part of the residual error is conditioned on scene state. The loop in Eq.~\ref{eq:state} is the central interface: Agents reason about scene evidence and memory, while the backbone continues to model temporal dynamics.

\subsection{Agent Roles and Communication}

Four Agent groups carry this communication path. External evidence Agents align asynchronous sources, remove metadata that could leak labels, and produce a fused environmental representation $z_t$. A frozen vision-language encoder is used inside this group for satellite, field-image, short-video, and report evidence. In our implementation this encoder is Qwen2.5-VL \citep{bai2025qwen25vl,radford2021clip,alayrac2022flamingo}, but the method only requires a visual-semantic interface. Forecasting, expert routing, and parameter approval remain assigned to downstream Agents. The encoder standardizes evidence for the Agent layer; numerical forecasts are produced later by the forecasting path.

Scene cognition and decision Agents compare $z_t$ with a known scene prototype memory. Their outputs are known-scene probabilities, unknown confidence, and the decision to use the known-scene residual path or pass a candidate to the evolution pathway. The routing Agent turns the accepted scene belief into expert weights. Let $K$ be the number of residual experts, $T_r$ the routing temperature, and $\eta$ the mixture coefficient between adaptive and uniform expert coverage:
\begin{equation}
 \begin{aligned}
 \pi_t&=\operatorname{softmax}(\operatorname{ROUTE}(z_t)/T_r),\\
 w_t&=\eta\pi_t+(1-\eta)\mathbf{1}/K,
 \end{aligned}
 \label{eq:routing}
\end{equation}
\begin{equation}
 r_t=\sum_{k=1}^{K}w_{tk}\operatorname{RESIDUAL}_k(p_t,m_t).
\end{equation}
We use hybrid routing because uniform routing protects against collapse but ignores scene evidence. Fully adaptive routing can over-concentrate on one expert.

At inference time, the system maintains an online residual acceptance Agent. For each window, the system logs both the backbone forecast and the Agent-corrected forecast. When the corresponding power observation later arrives, the Agent updates a rolling error ledger. Let $W_t$ be the most recent observed ledger window, and let $e_i^{(0)}$ and $e_i^{(A)}$ be the absolute errors of the backbone and Agent paths for entry $i$. The quantities $\rho_{\mathrm{mae}}$, $\rho_{\mathrm{rmse}}$, and $\rho_q$ are the relative reductions in MAE, RMSE, and the $q$-quantile absolute error over $W_t$. They are combined into a causal reliability score $g_t$, and the residual is accepted only when that score crosses a deployment threshold $\tau_g$:
\begin{flalign}
 &\begin{aligned}
 g_t&=(\rho_{\mathrm{mae}}+\rho_{\mathrm{rmse}}+0.5\rho_q)/2.5,\\
 \operatorname{GUARD}_t&=\mathbb{I}[g_t>\tau_g].
 \end{aligned}&&
 \label{eq:guard}
\end{flalign}
Here $\mathbb{I}[\cdot]$ is an indicator that returns 1 when the condition is true and 0 otherwise. In the reported runs, $\tau_g=0$, the ledger length is 2 observed entries, the minimum history is 2 entries, and $q=0.50$. A short ledger is the deployment setting because the guard must react after only a few observed windows, without waiting for a long calibration period. Sensitivity analyses evaluate ledger lengths 2, 4, 8, 16, and 32, threshold movement, and delayed feedback.

Candidate-scene Agents handle possible new regimes. Let $S_{\mathrm{known}}$ denote the known-scene set, $\mu_j$ a known-scene prototype, and $\tau_{\mathrm{nov}}$ the novelty threshold. A novelty Agent compares the fused scene representation with known prototypes and buffers candidate windows:
\begin{flalign}
 &\begin{aligned}
 s_t^{\mathrm{nov}}&=\min_{j\in S_{\mathrm{known}}}\{1-\operatorname{COS}(z_t,\mu_j)\},\\
 c_t&=\mathbb{I}[s_t^{\mathrm{nov}}>\tau_{\mathrm{nov}}].
 \end{aligned}&&
 \label{eq:novelty}
\end{flalign}
If enough buffered windows form a candidate cluster $c$, the parameter-evolution Agent initializes a virtual residual expert from nearest validated experts $N_c$. Existing residual experts are denoted by $\phi_j$, the candidate initialization by $\phi_c^{(0)}$, and the minimum validation gain by $\delta$:
\begin{equation}
 \phi_c^{(0)}=\sum_{j\in N_c}\alpha_j\phi_j.
 \label{eq:promotion}
\end{equation}
The backbone is frozen during this process; only residual/adaptation parameters are trained before a candidate is allowed to enter the pool. After numerical novelty checks, a long-reasoning language model writes the physical description and registry text for users. In our implementation, DeepSeek-v4 supplies this sidecar \citep{deepseekai2026v4,yao2023react,shinn2023reflexion}. Classification, routing, training, and promotion remain validation-controlled downstream steps. Registry entries provide audit information: they record which physical scene was observed and why a correction was retained, while the promotion decision itself remains validation-error based.

\begin{windalgorithm}{Agent-in-the-loop residual adaptation with candidate promotion}{alg:promotion}
\begin{algorithmic}[1]
\Require Causal packet $X_t$, memory $M_{t-1}$, expert pool $P_{t-1}$
\State $z_t\gets\operatorname{ENCODE}(X_t)$
\State $(s_t,s_t^{\mathrm{nov}})\gets\operatorname{SCENE}(z_t,M_{t-1})$
\If{$s_t^{\mathrm{nov}}\leq\tau_{\mathrm{nov}}$}
  \State $w_t\gets\operatorname{ROUTE}(s_t,P_{t-1})$
  \State $\tilde{y}_t\gets\hat{y}^{(0)}_t+\sum_k w_{tk}\operatorname{RESIDUAL}_k(\cdot)$
  \State \Return $\operatorname{GUARD}(\tilde{y}_t,\hat{y}^{(0)}_t,M_{t-1})$
\Else
  \State $c\gets\operatorname{BUFFER}(z_t,X_t)$
  \State $\phi_c^{(0)}\gets\operatorname{INTERPOLATE}(c,P_{t-1})$
  \State $\phi_c\gets\operatorname{TRAIN}(\phi_c^{(0)};\theta_B\text{ frozen})$
  \If{$\operatorname{VALERR}(\phi_c)\leq\operatorname{VALERR}(\phi_c^{(0)})-\delta$}
    \State $P_t\gets P_{t-1}\cup\{\phi_c\}$
  \EndIf
  \State $\operatorname{EXPLAIN}_{\mathrm{LLM}}(c)$ writes registry text only
\EndIf
\end{algorithmic}
\end{windalgorithm}

\subsection{Training and Evaluation Boundary}

All modalities are aligned causally: normalizers are fitted on training splits, timestamps and file names are excluded from model inputs, and visual metadata are not used as labels. Wind-power and SCADA inputs are motivated by public benchmarking resources \citep{iea2026task51,zhou2024sdwpf}; tropical cyclone context is matched from IBTrACS \citep{knapp2010ibtracs,gahtan2024ibtracs}; satellite quicklooks are treated as external observations from NASA GIBS \citep{nasa2026gibs}. The main endpoint compares backbone-only and backbone-plus-Agent prediction at leads 4, 8, 12, 24, and 48 under the same split and normalization protocol. The saved stream uses 30-minute sampling, so these leads correspond to 2, 4, 6, 12, and 24 hours. All main runs use seeds 42--46, sequence length 48, forecast head length 48, hidden dimension 128, batch size 64, four training epochs, and at most 12,000 training windows. Train-split statistics define normalization. Optimization uses AdamW with learning rate 0.003 and weight decay $10^{-4}$. The evaluated test stream contains 1,800 windows per lead from 1 October 2020 to 11 November 2020 in the saved prediction files. The released audit tables also record five-seed dispersion, per-run elapsed training time, and checkpoint size for each backbone and Agent pair.

\section{Experimental Design and Results}

\subsection{Experimental Design}

The forecasting endpoint uses the Spatial-Dynamic Wind Power Forecasting dataset, with public SDWPF power/SCADA records as the measured target source. Meteorological and event context are added from the project weather-event resources, IBTrACS tropical-cyclone records, and sanitized satellite quicklooks. All compared backbones use the same chronological train/validation/test split, the same input length, and the same forecast leads. File names, folder labels, and visual metadata are excluded from model inputs; the sanitized quicklook manifest is used only to locate images. Thus the Agent layer receives external scene evidence under the same causal boundary as the backbone receives temporal signals.

We organize the experiments around the Agent communication graph instead of a list of disconnected modules. Experiment 1 compares multi-horizon forecasts across the backbone zoo. Experiment 2 isolates scene-conditioned residual correction against a generic context vector. A final check examines whether known and candidate scene streams pass through the specified evidence boundary. Detailed local ablations are moved to the Appendix as audit material.

The evaluated leads are 4, 8, 12, 24, and 48. Metrics are MAE and RMSE reductions relative to paired controls; positive values indicate lower error after adding the Agent layer. Saved prediction files and routing outputs generate all figures, so reported claims can be traced to the deployed Agent path instead of detached summary tables. For the online-guard audit against generic selection rules, we replay the same paired residual forecasts through EMA, Hedge, dynamic model averaging, simple residual ensemble, random matched-coverage, matched-score, and oracle selectors. All selector baselines receive the same backbone and Agent predictions; only the acceptance decision changes.

\subsection{Multi-Horizon Forecasting}

\begin{figure*}[t]
  \centering
  \includegraphics[width=0.96\textwidth]{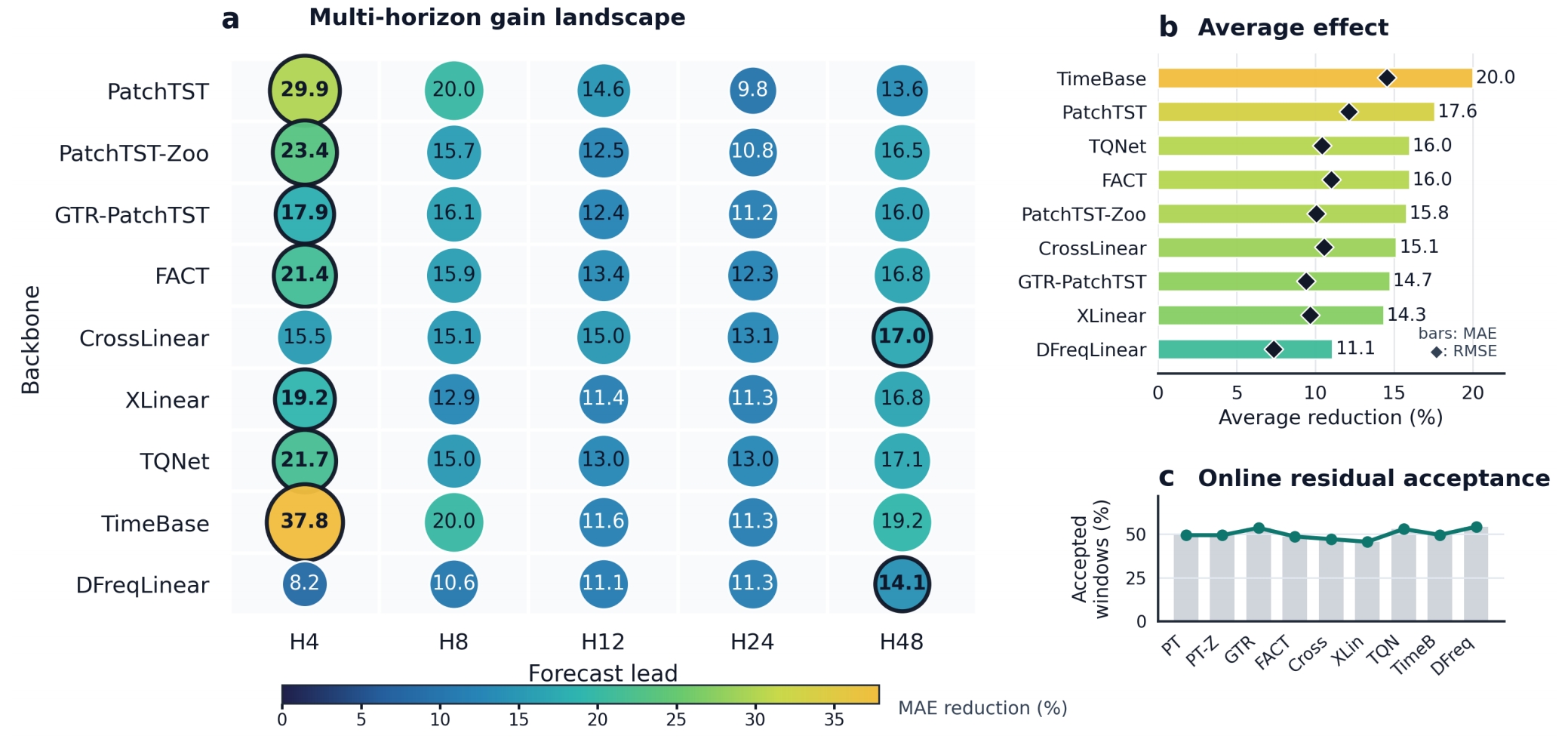}
  \caption{Multi-horizon Agent-in-the-loop gains at leads 4, 8, 12, 24, and 48. Panel (a) encodes paired MAE reduction by colour and circle area, with black rims marking the best lead for each backbone. Panel (b) compares average MAE and RMSE reductions, and panel (c) shows the online residual acceptance rate used by the guarded correction path.}
  \label{fig:multihorizon}
\end{figure*}

Table~\ref{tab:main-results} and Figure~\ref{fig:multihorizon} report lower errors after adding the Agent-in-the-loop layer to the evaluated time-series backbones. Averaged over leads 4, 8, 12, 24, and 48, MAE reductions range from 11.21\% for DFreqLinear to 20.26\% for TimeBase, and RMSE reductions range from 7.52\% to 14.97\%. The horizon-resolved matrix in Figure~\ref{fig:multihorizon} is the stricter check: every evaluated backbone-horizon pair has positive MAE and RMSE reduction under the same online acceptance controller. The size of the improvement depends on the forecasting architecture. Larger gains appear when the base forecaster leaves scene-sensitive residual structure, and smaller gains appear when the base forecaster already absorbs part of that structure. The pattern points to a specific mechanism: Agents help by deciding when scene-conditioned residual correction is justified, not by replacing temporal modelling. A moving-block bootstrap over saved windows gives an overall MAE reduction of 15.72\% with a 95\% interval of 15.44--15.99\%, and an RMSE reduction of 10.68\% with a 95\% interval of 10.46--10.94\%.

\begin{table*}[t]
\centering
\caption{Backbone-only and Agent-in-the-loop forecasts at all evaluated leads. Values are five-seed mean$\pm$std from the saved paired prediction protocol; reductions are computed per seed against the corresponding backbone-only control at the same lead and then averaged.}
\label{tab:main-results}
\scriptsize
\setlength{\tabcolsep}{2.8pt}
\renewcommand{\arraystretch}{0.90}
\resizebox{\textwidth}{!}{%
\begin{tabular}{llrrrrrr}
\toprule
Backbone & Lead & Base MAE & Agents MAE & MAE red. & Base RMSE & Agents RMSE & RMSE red. \\
\midrule
\multirow{5}{*}{PatchTST} & 4  & $202.20\pm26.44$ & $140.13\pm4.44$ & $29.89\pm7.88\%$ & $264.06\pm31.83$ & $206.34\pm9.60$ & $21.15\pm7.83\%$ \\
 & 8  & $231.83\pm14.78$ & $184.90\pm3.65$ & $19.95\pm5.93\%$ & $301.61\pm15.21$ & $262.93\pm10.08$ & $12.68\pm4.75\%$ \\
 & 12 & $249.87\pm8.54$ & $213.14\pm3.89$ & $14.61\pm3.63\%$ & $325.47\pm5.73$ & $296.00\pm3.72$ & $9.02\pm2.37\%$ \\
 & 24 & $258.48\pm5.56$ & $233.10\pm3.92$ & $9.81\pm0.89\%$ & $339.44\pm2.43$ & $316.46\pm1.81$ & $6.77\pm0.72\%$ \\
 & 48 & $270.68\pm5.20$ & $233.88\pm5.34$ & $13.58\pm2.13\%$ & $367.24\pm11.60$ & $327.02\pm5.24$ & $10.91\pm2.11\%$ \\
\midrule
\multirow{5}{*}{PatchTST-Zoo} & 4  & $183.62\pm8.40$ & $140.29\pm1.65$ & $23.44\pm4.14\%$ & $237.56\pm5.12$ & $204.50\pm3.55$ & $13.87\pm2.94\%$ \\
 & 8  & $220.64\pm9.73$ & $185.74\pm2.79$ & $15.66\pm4.48\%$ & $289.17\pm5.26$ & $263.06\pm7.77$ & $8.99\pm3.62\%$ \\
 & 12 & $241.12\pm7.14$ & $210.87\pm1.50$ & $12.48\pm2.77\%$ & $315.82\pm3.09$ & $292.64\pm5.27$ & $7.33\pm1.85\%$ \\
 & 24 & $255.78\pm3.23$ & $228.23\pm3.62$ & $10.77\pm0.84\%$ & $338.06\pm5.11$ & $312.52\pm5.95$ & $7.56\pm0.41\%$ \\
 & 48 & $277.66\pm4.90$ & $231.83\pm3.51$ & $16.49\pm1.46\%$ & $371.18\pm5.82$ & $324.54\pm2.43$ & $12.55\pm1.31\%$ \\
\midrule
\multirow{5}{*}{GTR-PatchTST} & 4  & $167.67\pm12.40$ & $137.13\pm2.23$ & $17.88\pm5.80\%$ & $225.55\pm11.82$ & $200.44\pm4.61$ & $10.95\pm4.89\%$ \\
 & 8  & $213.33\pm7.92$ & $178.70\pm4.62$ & $16.09\pm5.04\%$ & $280.05\pm7.13$ & $252.11\pm8.30$ & $9.91\pm4.20\%$ \\
 & 12 & $235.43\pm6.57$ & $206.25\pm7.05$ & $12.38\pm2.47\%$ & $313.15\pm2.88$ & $287.73\pm7.39$ & $8.12\pm2.15\%$ \\
 & 24 & $260.16\pm7.60$ & $230.81\pm5.13$ & $11.24\pm2.47\%$ & $335.94\pm1.77$ & $310.81\pm5.36$ & $7.48\pm1.50\%$ \\
 & 48 & $279.63\pm6.15$ & $234.92\pm6.08$ & $15.98\pm1.70\%$ & $361.45\pm4.42$ & $323.05\pm0.77$ & $10.61\pm1.16\%$ \\
\midrule
\multirow{5}{*}{FACT} & 4  & $176.96\pm19.12$ & $137.93\pm3.97$ & $21.39\pm7.93\%$ & $241.14\pm8.97$ & $206.70\pm8.55$ & $14.20\pm4.50\%$ \\
 & 8  & $210.22\pm14.56$ & $176.05\pm6.26$ & $15.87\pm7.27\%$ & $289.82\pm6.57$ & $259.98\pm17.77$ & $10.29\pm5.83\%$ \\
 & 12 & $232.95\pm18.05$ & $201.01\pm6.48$ & $13.37\pm6.09\%$ & $317.32\pm6.52$ & $288.65\pm13.18$ & $8.98\pm5.00\%$ \\
 & 24 & $250.70\pm12.95$ & $219.63\pm8.04$ & $12.33\pm1.68\%$ & $337.64\pm4.43$ & $307.64\pm2.04$ & $8.87\pm1.04\%$ \\
 & 48 & $272.26\pm9.04$ & $226.37\pm3.86$ & $16.82\pm1.40\%$ & $366.42\pm2.07$ & $319.90\pm4.61$ & $12.70\pm1.00\%$ \\
\midrule
\multirow{5}{*}{CrossLinear} & 4  & $157.76\pm4.61$ & $133.29\pm3.73$ & $15.48\pm2.66\%$ & $228.98\pm6.65$ & $204.72\pm4.22$ & $10.55\pm2.57\%$ \\
 & 8  & $199.72\pm8.36$ & $169.40\pm5.21$ & $15.14\pm1.61\%$ & $281.72\pm5.50$ & $255.09\pm3.19$ & $9.44\pm1.17\%$ \\
 & 12 & $228.29\pm11.41$ & $194.04\pm7.80$ & $14.97\pm1.36\%$ & $315.36\pm3.52$ & $283.37\pm4.08$ & $10.14\pm0.99\%$ \\
 & 24 & $249.73\pm13.74$ & $216.93\pm8.44$ & $13.05\pm2.56\%$ & $336.19\pm7.37$ & $304.07\pm5.18$ & $9.53\pm2.03\%$ \\
 & 48 & $272.15\pm5.90$ & $225.76\pm4.59$ & $17.03\pm1.57\%$ & $369.45\pm2.99$ & $321.12\pm4.22$ & $13.08\pm1.16\%$ \\
\midrule
\multirow{5}{*}{XLinear} & 4  & $166.01\pm2.54$ & $134.14\pm2.00$ & $19.20\pm0.78\%$ & $229.77\pm6.21$ & $202.40\pm3.59$ & $11.89\pm0.98\%$ \\
 & 8  & $201.69\pm4.31$ & $175.72\pm4.49$ & $12.87\pm1.63\%$ & $280.69\pm6.38$ & $258.99\pm2.73$ & $7.71\pm1.38\%$ \\
 & 12 & $225.48\pm3.20$ & $199.77\pm5.61$ & $11.41\pm1.98\%$ & $312.19\pm4.45$ & $289.12\pm2.40$ & $7.37\pm1.76\%$ \\
 & 24 & $246.60\pm9.71$ & $218.49\pm7.08$ & $11.35\pm2.63\%$ & $333.94\pm3.02$ & $306.84\pm5.30$ & $8.11\pm1.76\%$ \\
 & 48 & $268.29\pm10.52$ & $223.00\pm4.00$ & $16.82\pm2.33\%$ & $371.32\pm3.36$ & $321.83\pm4.46$ & $13.32\pm1.94\%$ \\
\midrule
\multirow{5}{*}{TQNet} & 4  & $174.72\pm14.96$ & $135.96\pm2.42$ & $21.69\pm7.31\%$ & $230.36\pm20.20$ & $199.26\pm6.98$ & $13.09\pm6.64\%$ \\
 & 8  & $212.60\pm11.98$ & $180.19\pm2.73$ & $15.02\pm5.08\%$ & $280.48\pm13.43$ & $255.01\pm8.06$ & $8.97\pm3.75\%$ \\
 & 12 & $239.45\pm8.73$ & $208.15\pm1.10$ & $12.97\pm3.39\%$ & $314.54\pm5.74$ & $289.80\pm4.98$ & $7.85\pm1.93\%$ \\
 & 24 & $257.66\pm3.44$ & $224.08\pm3.02$ & $13.03\pm0.20\%$ & $336.31\pm2.84$ & $305.87\pm2.13$ & $9.05\pm0.32\%$ \\
 & 48 & $280.98\pm2.69$ & $232.84\pm4.94$ & $17.13\pm1.91\%$ & $370.20\pm6.01$ & $321.82\pm2.86$ & $13.05\pm1.74\%$ \\
\midrule
\multirow{5}{*}{TimeBase} & 4  & $236.20\pm3.77$ & $146.82\pm0.65$ & $37.83\pm1.18\%$ & $312.71\pm1.42$ & $217.42\pm1.44$ & $30.47\pm0.47\%$ \\
 & 8  & $243.92\pm4.18$ & $195.20\pm0.67$ & $19.96\pm1.57\%$ & $324.04\pm0.80$ & $278.69\pm1.62$ & $13.99\pm0.51\%$ \\
 & 12 & $250.12\pm4.52$ & $221.04\pm1.49$ & $11.61\pm1.17\%$ & $332.60\pm0.81$ & $307.80\pm1.09$ & $7.46\pm0.43\%$ \\
 & 24 & $259.06\pm5.15$ & $229.71\pm3.58$ & $11.32\pm0.90\%$ & $345.29\pm1.66$ & $318.42\pm1.18$ & $7.78\pm0.66\%$ \\
 & 48 & $274.35\pm6.62$ & $221.56\pm2.16$ & $19.22\pm1.43\%$ & $366.36\pm2.09$ & $318.99\pm1.39$ & $12.93\pm0.75\%$ \\
\midrule
\multirow{5}{*}{DFreqLinear} & 4  & $143.42\pm4.32$ & $131.55\pm1.42$ & $8.23\pm2.24\%$ & $211.18\pm2.60$ & $199.94\pm3.97$ & $5.33\pm1.06\%$ \\
 & 8  & $199.56\pm3.94$ & $178.37\pm3.86$ & $10.61\pm1.69\%$ & $274.08\pm2.64$ & $256.18\pm6.23$ & $6.54\pm1.47\%$ \\
 & 12 & $232.45\pm5.03$ & $206.65\pm3.50$ & $11.08\pm1.85\%$ & $311.59\pm2.29$ & $289.79\pm5.04$ & $7.00\pm1.26\%$ \\
 & 24 & $254.91\pm6.04$ & $225.96\pm5.33$ & $11.34\pm1.45\%$ & $335.10\pm1.45$ & $308.77\pm2.52$ & $7.86\pm0.89\%$ \\
 & 48 & $278.09\pm1.49$ & $238.80\pm4.91$ & $14.13\pm1.45\%$ & $359.71\pm4.31$ & $323.52\pm2.19$ & $10.05\pm1.25\%$ \\
\bottomrule
\end{tabular}}
\end{table*}

\begin{figure*}[t]
  \centering
  \includegraphics[width=0.92\textwidth]{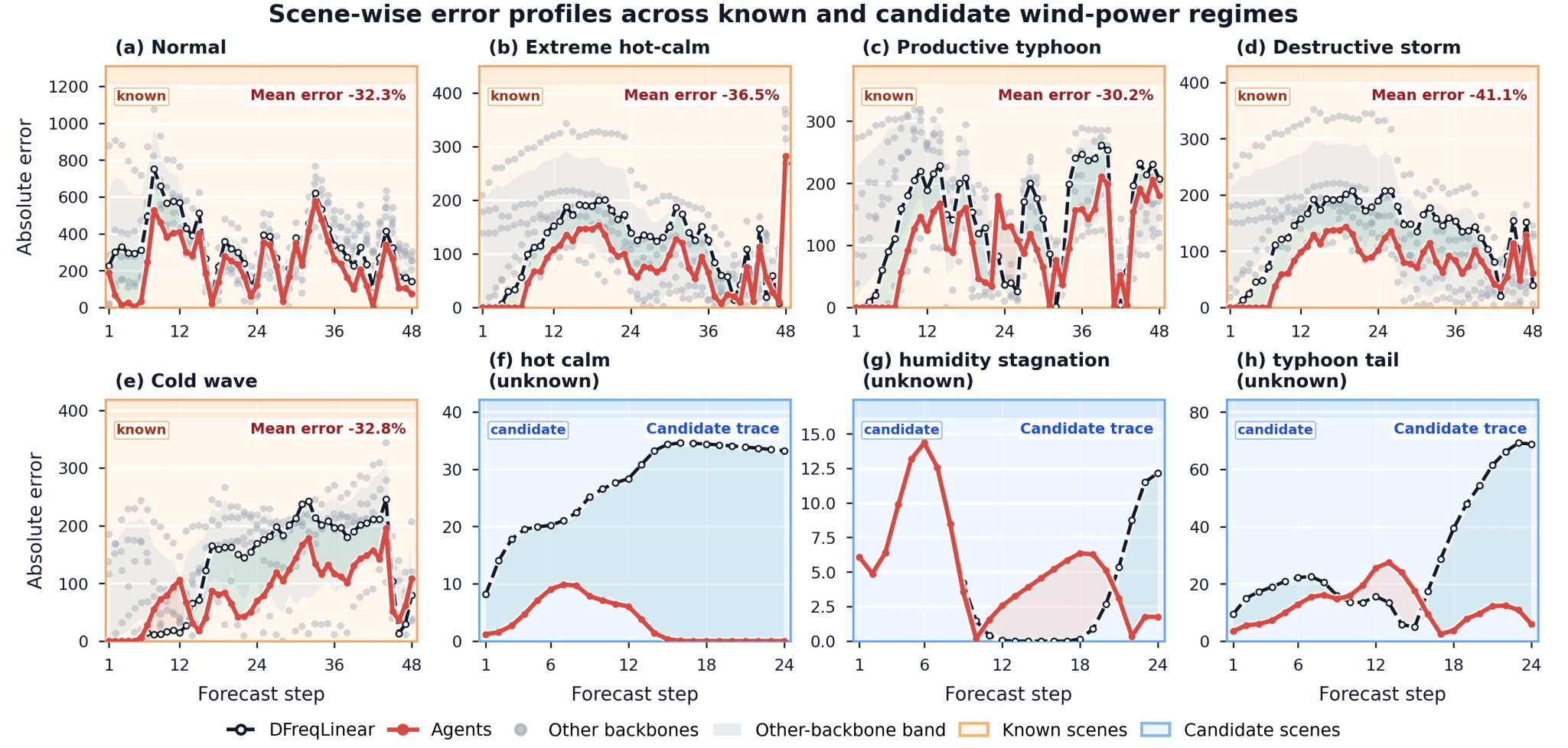}
  \caption{Scene-wise error trends for known and candidate wind-power regimes. Orange panels mark known scenes and blue panels mark candidate-scene streams. The $y$-axis is absolute forecast error, so no observed-power curve is drawn. DFreqLinear is shown as the reference backbone trace, Agents as the guarded residual path, grey points denote other backbone-only controls, and the grey ribbon summarizes their error range.}
  \label{fig:scene-errors}
\end{figure*}

\subsection{Mechanism Diagnostics}

The residual-pool ablation comes first because it is the shortest path from scene evidence to forecast correction. Under the saved fixed protocol, the backbone-only control gives MAE 58.188 and RMSE 95.021, while the backbone plus Agents gives MAE 54.986 and RMSE 91.853, corresponding to 5.50\% MAE and 3.33\% RMSE reductions. This unguarded check shows that the residual pool supplies a usable correction path, but Table~\ref{tab:ablation} attributes the larger deployed gain mainly to the online guard. Routing diagnostics further explain why the residual path must be structured. Uniform routing spreads mass over experts but ignores scene evidence; fully adaptive routing may collapse. Hybrid routing gives the lowest saved MAE/RMSE in the strict diagnostic while keeping several experts active. The Appendix contains the full mechanism dashboard and known/candidate scene diagnostic as audit views instead of additional primary endpoints.

\begin{table}[t]
\centering
\caption{Component ablation for the Agent-in-the-loop residual path. The rows separate residual experts, mixture-of-experts routing, and the online guard, showing which part of the Agent layer contributes to the reported gains.}
\label{tab:ablation}
\scriptsize
\setlength{\tabcolsep}{2.5pt}
\renewcommand{\arraystretch}{1.5}
\resizebox{\columnwidth}{!}{%
\begin{tabular}{p{0.50\columnwidth}ccccc}
\toprule
Variant & Pool & MoE & Guard & MAE red. & RMSE red. \\
\midrule
Backbone only & -- & -- & -- & 0.00\% & 0.00\% \\
Residual pool + static expert & Y & static & -- & 5.28\% & 3.27\% \\
Residual pool + uniform experts & Y & uniform & -- & 5.33\% & 3.25\% \\
Residual pool + soft adaptive MoE & Y & soft & -- & 5.44\% & 3.03\% \\
Residual pool + hybrid MoE & Y & hybrid & -- & 5.50\% & 3.33\% \\
Residual pool + MoE + online guard & Y & hybrid/learned & Y & 15.57\% & 10.59\% \\
Online guard contribution over unguarded Agent & -- & same & Y & 15.06\% & 11.96\% \\
\bottomrule
\end{tabular}}
\end{table}

\begin{figure}[b]
  \centering
  \includegraphics[width=\columnwidth,height=0.28\textheight,keepaspectratio]{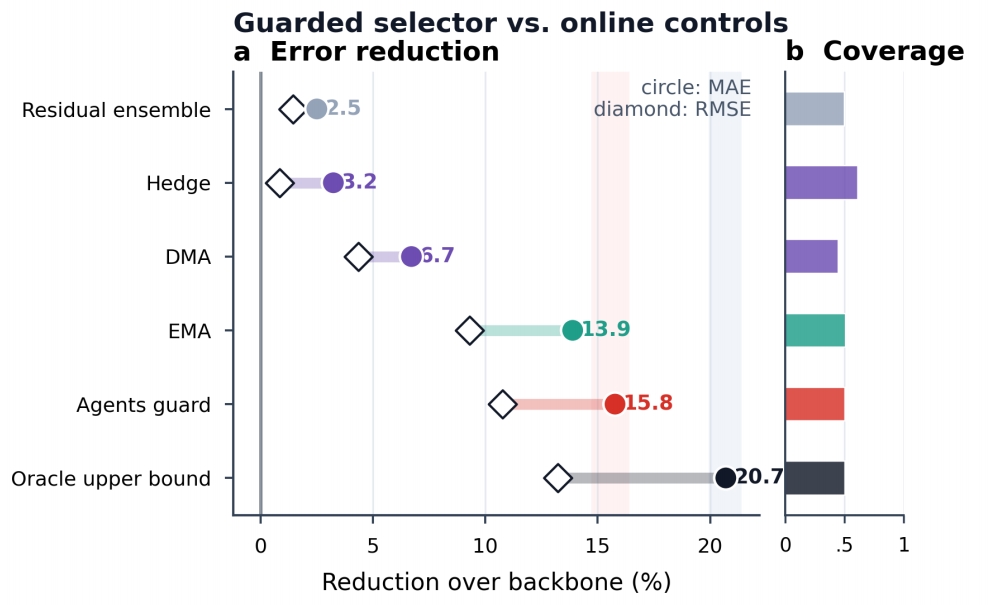}
  \caption{Main-text selector comparison. Circles show MAE reduction, diamonds show RMSE reduction, and the aligned right panel shows residual acceptance coverage under the same saved paired forecasts. The full coverage-risk and ledger diagnostics are in Appendix Figure 5.}
  \label{fig:selector-comparison}
\end{figure}

For the online acceptance rule, Table~\ref{tab:selector-baselines} compares direct model-selection baselines under the same saved backbone and raw-Agent predictions for every row. Raw Agent correction and random matched coverage fail to control RMSE, which shows why acceptance is needed. The Agent guard remains causal and gives the largest non-oracle reductions among the reported selectors under comparable saved forecasts. The table is an online-selection audit, not direct evidence of scene reasoning. The oracle row is only a non-causal upper bound. Accepted windows have lower Agent-path error than backbone-path error; rejected windows show the reverse pattern. The guard is designed to capture precisely this risk structure. Residual coverage is also horizon-dependent: it is 0.58, 0.50, 0.47, 0.47, and 0.50 at leads 4, 8, 12, 24, and 48, respectively. The ``Max reject run'' column reports the longest uninterrupted sequence of windows for which a selector rejects the residual path and uses the backbone forecast instead; it is included to expose whether a method remains responsive or stays inactive for long stretches. Figure~\ref{fig:selector-comparison} keeps only the central selector comparison in the main text; the coverage-risk, ledger, and accepted/rejected-window panels are reported in the Appendix.

\begin{table}[t]
\centering
\caption{Direct online-selection baselines on the same saved residual forecasts. Coverage is the fraction of windows where the residual path is used; DMA coverage is its mean residual weight. Max reject run is the longest consecutive run of forecast windows in which the selector rejects the residual path and falls back to the backbone, averaged over backbone--seed--lead cells. The fixed residual ensemble is not a binary accept/reject selector, so its reject-run value is not interpreted. Oracle uses current-window labels and is reported only as an upper bound.}
\label{tab:selector-baselines}
\scriptsize
\setlength{\tabcolsep}{2.6pt}
\renewcommand{\arraystretch}{1.35}
\resizebox{\columnwidth}{!}{%
\begin{tabular}{lrrrr}
\toprule
Selector & Coverage & MAE red. & RMSE red. & Max reject run \\
\midrule
Backbone only & 0.00 & 0.00\% & 0.00\% & 1800.0 \\
Raw Agent & 1.00 & 0.65\% & 0.55 & 0.0 \\
Residual ensemble & 0.50 & 2.51\% & 1.45\% & 1800.0 \\
Random matched & 0.46 & 0.96\% & 0.78 & 11.8 \\
Hedge & 0.61 & 3.24\% & 0.86\% & 507.4 \\
DMA & 0.45 & 6.72\% & 4.35\% & 330.9 \\
EMA selector & 0.51 & 13.89\% & 9.31\% & 70.8 \\
Agents guard & 0.50 & 15.77\% & 10.77\% & 55.9 \\
Oracle upper bound & 0.51 & 20.70\% & 13.22\% & 53.1 \\
\bottomrule
\end{tabular}}
\end{table}

\section{Discussion and Limitations}

The empirical pattern matches a specific mechanism. Agent-in-the-loop forecasting is useful when scene reasoning is expressed as guarded residual correction, not as a new temporal backbone. Performance improvements vary across forecasting architectures because each backbone leaves a different residual pattern. Early/late quicklook reductions of 15.80\% and 15.42\% check temporal concentration, and candidate-scene runs audit promotion. Long-term unattended evolution would require independent wind farms, event-disjoint cross-region validation, larger rare-event support, and latency/parameter analyses. Qwen2.5-VL and DeepSeek-v4 are restricted to scene-evidence encoding and explanation; numerical Agents handle forecasting, routing, training, and promotion.

Post-event review benefits from the same decomposition. When a forecast fails, the logged path shows whether the error came from the temporal backbone, the scene belief, the selected residual expert, or the acceptance decision. Wind-farm operators can then inspect a small number of Agent choices without replacing the backbone that already fits existing forecasting practice. Rejected corrections remain available for diagnosis, but they do not alter the delivered forecast. The log also makes the benchmark easier to reproduce. Each delivered forecast has a base value, a candidate correction, a routing vector, and a guard decision. Applying the stored guard to the paired predictions recovers the final output. The appendix reports selector and ledger diagnostics alongside aggregate error.

\clearpage
\bibliography{references}
\end{document}